\documentstyle[aps,multicol,prl]{revtex}
\begin{document}
%\draft
\title{ Field theory for charged fluids and colloids}

\author{
Roland R. Netz$^{\$}$ and 
Henri Orland$^*$
}
\address{$^{\$}$Max-Planck-Institut f\"ur Kolloid- 
und Grenzfl\"achenforschung,
Kantstr. 55, 14513 Teltow, Germany}
\address{$^*$Service de Physique Th\'eorique, 
CEA-Saclay, 91191 Gif sur Yvette, France}

\date{\today}
\maketitle

\begin{abstract}
A systematic field theory is presented for charged systems.
The one-loop level corresponds to the classical Debye-H\"uckel (DH) theory,
and exhibits the full hierarchy of multi-body correlations determined by
pair-distribution functions given by the screened DH
potential. Higher-loop corrections 
can lead to attractive pair interactions between 
colloids in asymmetric ionic environments.
The free energy follows as a loop-wise  expansion in half-integer 
powers of the density;
the resulting two-phase demixing region
shows pronounced  deviations from DH theory
for strongly charged colloids.
 \end{abstract}

\begin{multicols}{2} 

Since the early work of Debye and H\"uckel (DH) it is known 
that electrostatic interactions in a
mixture of positively and negatively charged particles
produce a net attraction\cite{DH}. 
This is due to charge screening:
Each charge is (on average) predominantly surrounded by oppositely
charged particles, which thus leads to an overall attraction
between the particles. The resulting DH free energy contribution
 has been theoretically
demonstrated to lead to phase separation in the context
of ionic fluids\cite{Fisher1} and colloidal mixtures\cite{Roij},
in agreement with experimental\cite{Sengers} and
numerical\cite{Valleau} work on ionic fluids and experiments
on colloidal mixtures\cite{Tata}.
The exact nature and origin of the DH term has remained somewhat 
unclear, and several improvements have been devised based on 
series expansions and liquid-state theory\cite{Stell},
explicit incorporation of dipole pairs\cite{Fisher1}, and
density-functional theory\cite{Roij}.

A second intensely debated question concerns the
possible existence of attractive interactions between
similarly charged  objects in electrolyte solution\cite{Jensen}.
Experimentally,
such an attraction has been seen  for 
DNA\cite{Bloomfield}
and strongly charged microspheres which are 
confined between charged plates\cite{Crocker}. 
Clearly, the phase separation observed
for colloidal mixtures\cite{Tata} is a priori {\em not} an 
indication for such an 
attractive interaction, because the dense phase is induced by
attractions between oppositely charged particles, as becomes explicit
within DH theory (also, see the discussion in \cite{Roij}).

In this article we present a systematic field theory for charged 
systems, and calculate
both the free energy and
the effective interactions between charged particles 
immersed in an electrolyte solution. 
At the one-loop level, we recover the classical DH theory,
the nature of which transpires in an especially lucid
fashion within our framework:
we find the full hierarchy of multi-body correlations 
to be present, with 
all pair-distribution functions given by the screened
DH interaction.  This means in specific that 
triplet correlations are already included at the DH level
(in contrast to implicit assumptions in recent theories\cite{Fisher1}),
and that effective interactions between similarly charged particles are
repulsive.
At higher order in our theory (which corresponds to including 
multi-loop diagrams), non-trivial multi-body interactions
appear, and, consequently, the multibody correlations acquire
contributions which {\em cannot} be described as superposition
of pair correlations. 
Also, the effective pair interaction receives corrections which can
be attractive if i) the electrolyte
is asymmetric and consists of multivalent counterions and 
monovalent coions, or if ii) the colloidal charge is overcompensated
by salt ions.
The latter situation is realized in experiments on charged microspheres,
where a strong attraction is only found in the vicinity of a charged
wall\cite{Crocker}. 
The free energy of an ionic solution, expanded in the number of loops,
follows to be a series in half-integer powers of the density,
and thus constitutes a systematic low-density expansion\cite{McQuarrie}.
The effects of higher-loop contributions on the demixing transition
become increasingly important for highly charged colloids and lead to
pronounced deviations from DH theory.

To proceed, we consider the 
partition function of  $N$ charged, fixed  test particles, 
immersed in a multi-component electrolyte solution
with (in general) $M$ different types of ions,
\begin{eqnarray}
\label{part1}
Z[\{R_N\}]  &=&
\prod_{j=1}^M \left[ \frac{1}{n_j !}
\prod_{k=1}^{n_j} 
\int \frac{{\rm d} {\bf r}_k^{(j)}}{\lambda^3} \right]
\nonumber \\ && \exp \left\{ -
\frac{1}{2} \int {\rm d} {\bf r}
 {\rm d} {\bf r}' \hat{\rho}_c({\bf r})
v({\bf r}-{\bf r}')  \hat{\rho}_c({\bf r}') \right\},
\end{eqnarray}
where $v({\bf r}) = \ell_B /r$ is the Coulomb operator
and the charge density operator $\hat{\rho}_c$ is defined by
\begin{equation}
\hat{\rho}_c({\bf r}) \equiv  \sum_{i=1}^N Q_i\delta({\bf r} -{\bf R}_i)
+\sum_{j=1}^M \sum_{k=1}^{n_j} q_j\delta({\bf r} -{\bf r}_k^{(j)} )
\end{equation}
with $Q_i$ and $q_j$ being the charges (in units of the elementary 
charge $e$) of the test particles and the ions, respectively.
The length $\lambda$ is an arbitrary constant, and
the Bjerrum length $\ell_B \equiv e^2/4 \pi \epsilon k_B T$ defines
the length at which two unit charges interact with thermal
energy $k_BT$.
Electroneutrality of course requires 
$\sum_{i=1}^N Q_i  +\sum_{j=1}^M n_j q_j =0$.
\end{multicols}
Noting that the inverse Coulomb operator can be explicitly written as
$v^{-1}({\bf r}) = - \nabla^2 \delta({\bf r}) / 4 \pi \ell_B $,
after a Hubbard-Stratonovich transformation,
the partition function is given by
\begin{equation}
\label{part1b}
Z[\{R_N\}] = \int \frac{{\cal D}\phi}{Z_0} 
\exp \left\{-\frac{1}{8 \pi \ell_B}
\int {\rm d}{\bf r}(\nabla \phi)^2 - i 
\sum_{i=1}^N Q_i \phi({\bf R}_i) +
\sum_{j=1}^M n_j \log \left[\int \frac{{\rm d}{\bf r} }{V}
{\rm e}^{- i q_j \phi({\bf r}) } \right] +{\cal S} \right\},
\end{equation}
where $Z_0$ is
the partition function of the inverse Coulomb operator,
$Z_0 \sim \det v$,
and the entropy of ideal mixing is 
${\cal S} \equiv - \sum_j n_j \ln (\lambda^3 c_j)$ with 
$c_j \equiv n_j/V$ denoting the concentration of ion species $j$.
Performing a cumulant expansion of  (\ref{part1b}) in powers of
$\phi$, we can rewrite the partition function as 
\begin{equation}
\label{part2}
Z[\{R_N\}] = \int \frac{{\cal D}\phi}{Z_0} \exp \left\{-\frac{1}{2 }
\int {\rm d}{\bf r}{\rm d}{\bf r}'
\phi({\bf r}) v^{-1}_{\rm DH}({\bf r}-{\bf r}') \phi({\bf r}')
- i \sum_{i=1}^N Q_i \phi({\bf R}_i) +
W[\phi] +{\cal S} \right\},
\end{equation}
where $v_{\rm DH}$ is determined via
the inverse operator equation (the so-called Dyson equation in field theory)
\begin{equation}
\label{defDH}
v^{-1}_{\rm DH}({\bf r}) \equiv v^{-1} ({\bf r}) + I_2 \delta ({\bf r})
\end{equation}
which is solved by the well-known DH interaction
$ v_{\rm DH}({\bf r}) = \ell_B e^{-r \kappa}/r $
with  the screening length $\kappa^{-1}$ defined by
$\kappa^2 \equiv 4 \pi \ell_B I_2$.
All anharmonic terms are contained in the non-local potential $W$, which
is up to eighth order given by
\begin{eqnarray}
\label{W}
W[\phi] &=&  \frac{i I_3 V}{3!} \overline{\phi^3} +
\frac{I_4 V}{4!} \left(\overline{\phi^4}-3\overline{\phi^2}^2 \right)
-\frac{i I_5 V}{5!}  \left(\overline{\phi^5}-10 \overline{\phi^2}
\; \overline{\phi^3} \right) -
\frac{I_6 V}{6!} \left(\overline{\phi^6}-15 \overline{\phi^4} \;
\overline{\phi^2}-10  \overline{\phi^3}^2+30\overline{\phi^2}^3\right) 
\nonumber \\ && +
\frac{I_8 V}{8!} \left(\overline{\phi^8}-28 \overline{\phi^6}\;
\overline{\phi^2}-56 \overline{\phi^5}\;\overline{\phi^3}-35
\overline{\phi^4}^2+420\overline{\phi^4}\; \overline{\phi^2}^2+
560\overline{\phi^2}\; \overline{\phi^3}^2-630
\overline{\phi^2}^4\right).
\end{eqnarray}
We have introduced the  generalized ionic strength $I_n$, 
which is defined as 
$ I_n \equiv \sum_{j=1} q_j^n c_j $ 
and can take both positive and negative values.
In these equations $\overline{\phi^n}$ denotes moments of the field,
$
\overline{\phi^n } \equiv \int {\rm d}{\bf r} \phi^n({\bf r})/V$. 
The action is invariant with respect to a change
of the gauge field $\overline{\phi}$. We therefore set $\overline{\phi}=0$.
The linear term in $\phi$ in Eq.(\ref{part2}) can be removed by a shift
of the fluctuating field $\phi$, and the 
partition function then takes the form
\begin{equation}
\label{part3}
Z[\{R_N\}] =  \exp \left\{ {\cal S}- \frac{1}{2 } \sum_{i,j} Q_i Q_j
v_{\rm DH}({\bf R}_i-{\bf R}_j) \right\}
\int \frac{{\cal D}\phi}{Z_0} \exp \left\{-\frac{1}{2 }
\int {\rm d}{\bf r}{\rm d}{\bf r}'
\phi({\bf r}) v^{-1}_{\rm DH}({\bf r}-{\bf r}') \phi({\bf r}')
+ W[\tilde{\phi} ]  \right\}
\end{equation}
\begin{multicols}{2}
\narrowtext
%\noindent
where
$\tilde{\phi}({\bf r}) \equiv \phi({\bf r})
-i \sum_{i} Q_i v_{\rm DH}({\bf r}-{\bf R}_i)$.
Up to this point, our calculations are (in principle) {\em exact}:
keeping  terms of all powers in $W$ in (\ref{part3})
leads to a model equivalent
to the original partition function (\ref{part1}). They are
also {\em systematic}, in that keeping terms of higher and higher
order of $W$ should make the resulting theory a more and more
faithful representation of the underlying physical model. 
In fact, we will demonstrate that the lowest approximation, 
where the potential $W$ and thus anharmonic terms in $\phi$
are neglected altogether,
is equivalent to the classical DH theory. In this case, 
it follows from  (\ref{part3}) that 
the dimensionless pair interaction $U_2$ between
two test particles is just the DH potential, 
\begin{equation} \label{U2}
U_2({\bf R}_1-{\bf R}_2)  = Q_i Q_j v_{\rm DH}({\bf R}_1-{\bf R}_2) 
\end{equation}
and the two-point correlation function is 
$g_2({\bf R}_1-{\bf R}_2) \propto e^{-U({\bf R}_1-{\bf R}_2)} $
with a proportionality constant such that it
is normalized\cite{comment1}. Neglecting $W$,
there are no multibody
interactions between test particles in (\ref{part3}), and 
higher-order correlation functions are therefore given by 
products of the pair correlation function, $ 
g_3({\bf R}_1,{\bf R}_2,{\bf R}_3) \propto g_2({\bf R}_1-{\bf R}_2)
g_2({\bf R}_2-{\bf R}_3) g_2({\bf R}_1-{\bf R}_3)$,
and so on. This is the superposition principle, known as a postulate
from liquid state theory; it is exactly obeyed in DH theory.
To connect to liquid state theory, we note that
from Eq.(\ref{defDH}) one obtains
by inversion the integral equation (which in fact
holds for {\em any} pair interaction $v$)
\[
v({\bf r}) = v_{\rm DH}({\bf r}) + I_2 \int {\rm d} {\bf r}'
v_{\rm DH}({\bf r}') v({\bf r}-{\bf r}'),\]
the field-theoretic version of the Ornstein-Zernicke equation.
The DH interaction is the exact 
solution of this integral equation\cite{comment2}. 
The DH theory therefore 
contains correlations of all orders, and there is no need to 
explicitly add higher-order correlations (compare \cite{Fisher1}).
Improvements
can only come from adding non-trivial higher-body effective
interactions, i.e., from violations of the superposition principle.
This is the effect of the potential $W$, as we will 
demonstrate in the following.
\end{multicols}
\widetext
Expanding $W[\phi]$ in the exponential of Eq.(\ref{part3}), 
the first correction
comes from the cubic term,
\begin{equation}
\label{Z3}
Z[\{R_N\}] \propto   \exp \left\{ {\cal S}- \frac{1}{2 } \sum_{i,j} Q_i Q_j
v_{\rm DH}({\bf R}_i-{\bf R}_j) 
-{ I_3 \over 6} \sum_{i,j,k} Q_i Q_j Q_k 
\Omega_3({\bf R}_i,{\bf R}_j,{\bf R}_k)
\right\}
\end{equation}
with the three-point vertex given by
%\begin{equation}
$
\Omega_3({\bf R}_1,{\bf R}_2,{\bf R}_3) \equiv
\int{\rm d} {\bf r} \; v_{\rm DH}({\bf r}-{\bf R}_1)
v_{\rm DH}({\bf r}-{\bf R}_2) v_{\rm DH}({\bf r}-{\bf R}_3).
$
%\end{equation}
\begin{multicols}{2}
\narrowtext
The summation over $(i,j,k)$ in (\ref{Z3}) is unrestricted.
By considering the case where two of the three indices are equal, one obtains
a correction to the pair interaction, which reads
\begin{equation} \label{DU2}
\Delta U_2({\bf R}) =\frac{I_3 \ell_B^3 (Q_1^2 Q_2+Q_1 Q_2^2)}{6}
\; \Xi(R \kappa),
\end{equation}
where the function $\Xi$ (which is positive)
 is determined by the integral
$\Xi(x) \equiv  \int {\rm d}{\bf r} e^{-2r}
e^{-|{\bf r}-{\bf x}|}/ r^2
|{\bf r}-{\bf x}| = 
2\pi\left\{ e^{-x}\left( \ln 3-\Gamma[0,x]\right) 
+e^x \Gamma[0,3x] \right\}/x. $
The asymptotic behavior of $\Xi$  is
\begin{equation}
\Xi(x) \simeq  
     \left\{ \begin{array}{llll}
     &  -4 \pi \ln x 
         & {\rm for} &  x \ll 1  \\
     & 2\pi \ln 3 \frac{\displaystyle  e^{-x}}{\displaystyle x}
         & {\rm for }     & x \gg 1, \\
                \end{array} \right.
\end{equation}
and thus shows the same asymptotic behavior as the DH repulsion for
large separations.
When do we expect attractive interactions between likely charged particles? 
In other words, when does the prefactor in 
Eq.(\ref{DU2}) become
negative? If we assume the test particles to be positively charged, the 
condition for attraction is that the third-order ionic strength $I_3$ 
is negative.
Assuming a homogeneous {\em solution} of positive 
macroions with charge $Z$, concentration $c$,  and 
counter ions of valency $z$, the third-order ionic strength is 
$I_3 = c Z(Z^2-z^2)$ and clearly always positive: similarly
charged particles at finite concentration do not attract each other,
in agreement with experiments. 
 On the other hand, 
considering two {\em single}  charged macroions, 
it is easy to see that $I_3$
is negative if the salt solution is asymmetric:
for positive macroions, attraction is therefore 
possible i) if one has an electrolyte
consisting of negative ions with a higher valency than the positive ions
or ii) if there are more negative than positive ions in the local environment
(as in experiments between two charged walls\cite{Crocker}).
Comparing the strength of the DH repulsion Eq.(\ref{U2})
and the attraction Eq.(\ref{DU2}) at large separation we find the
attraction to dominate for $c \ell_B^3 Z^2>9(m^2+1)/(m^3-1)^2 \pi \ln^23$ 
for the case of a $m:1$ electrolyte\cite{comment3}. 
Experimentally, it is well-known that
asymmetric salts like Calciumchloride
induce the precipitation
of negatively charged macroions 
and negatively charged polymers\cite{Bloomfield}.

The phase behavior of charged colloidal mixtures follows from the
free energy with all particle coordinates integrated over,
\begin{equation} \label{free1}
{\cal F} = -\ln Z = -{\cal S} -\ln \left[ \frac{Z_2}{Z_0} \right]
- \ln \left\langle e^{W[\phi]}\right\rangle.
\end{equation}
The DH partition function is $Z_2 \sim \det v_{\rm DH} $,
whereas  higher-order correlations are contained in $W$. The expectation
value in (\ref{free1}) is evaluated with the DH propagator.
We first evaluate the DH free energy,
\[
f_{\rm DH} \equiv
-\frac{a^3}{V} \log \left[ \frac{Z_2}{Z_0} \right] =
-\left(\frac{a}{2 \pi}\right)^3 \int {\rm d} {\bf q} \log \sqrt{
\frac{ q^2}{q^2 + \kappa^2}}
\]
where the momentum integral goes over a cube of length $2 \pi/a$.
Since the integrand is isotropic we distort the integration
volume to a sphere and  obtain
 after a straightforward integration
\[
f_{\rm DH}  = -
\frac{a^3 \kappa^3}{6 \pi^2 } \arctan\left[\frac{\pi }{a \kappa}\right] +
\frac{a^2 \kappa^2}{6  \pi} +
\frac{ \pi}{12} \log\left [1+\frac{a^2 \kappa^2}{\pi^2}\right].
\]
In the limit $a \rightarrow 0$ one obtains the well-known result
$f _{\rm DH} \simeq -a^3\kappa^3 /12 \pi$ 
(plus corrections which scale linearly in
$\kappa^2$ and thus correspond to an unimportant shift in the chemical
potential). Our corrections as a function of the cut-off 
take in an approximate fashion the finite ion-sizes into account.
The free energy contribution 
$\Delta f \equiv -\frac{a^3}{V} \ln \left\langle e^{W[\phi]}\right\rangle$
is, using Eq.(\ref{W}), given by
\[
\Delta f 
= \frac{a^3 I_3^2}{12}\left[ \chi_3 + \frac{3}{2}  \langle \phi^2 \rangle^2
\chi_1\right] -\frac{a^3 I_4^2}{48} \chi_4 +
\frac{ a^3 I_5 I_3}{32} \langle \phi^2 \rangle^3 \chi_1.
\]
The expectation value $\langle \phi^2 \rangle$ is 
\[ \langle \phi^2 \rangle = \frac{\ell_B}{2 \pi^2} \int
\frac{{\rm d}{\bf q}}{q^2+\kappa^2}=\frac{2 \ell_B}{a}\left[
1-\frac{a \kappa}{\pi} \arctan\left(\frac{\pi}{a \kappa}\right)\right]
\]
and the generalized susceptibility $\chi_n$ is defined as
\[
\chi_n \equiv \int {\rm d}{\bf r} \langle \phi_0 \phi_r \rangle^n =
4 \pi \ell_B^n (n \kappa)^{n-3} \Gamma[3-n, an \kappa].
\]
Naive scaling predicts the generalized susceptibilities in $\Delta f$ to 
scale like $\chi_n \sim c^{(n-3)/2}$ as a function of the
ion density $c$. Since the dominant  terms in $\Delta f$ scale
as $I_n^2 \chi_n$, one would
thus obtain a systematic free-energy
 expansion in half-integer powers of the density, 
starting with the DH term $f_{\rm DH}$, 
which asymptotically scales as $c^{3/2}$\cite{McQuarrie}. 
In practice, the integrals in  $\chi_n$
diverge in the ultraviolet and thus depend on the
ion radius $a$ in a crucial way, which leads to changes from the naive 
scaling picture. 
We now present results for the simplified case of colloids of charge $Z$ and
concentration $c$ with counterions of valency $z$. Introducing the
energy scale $\epsilon \equiv  zZ \ell_B/a$ and the 
total ion volume fraction  $\tilde{c} \equiv a^3 c(1+Z/z)$ the DH theory
(which amounts to neglecting the term $\Delta f$)
predicts a critical point at $\epsilon \simeq 5.63$ and $\tilde{c} \simeq 
0.0418$, independent of the colloid charge $Z$. 
Including higher-order terms  contained in 
$\Delta f$ the critical interaction
strength $ \epsilon$ and volume fraction $\tilde{c}$ depend on $Z/z$,
as shown in Fig. 1. The limiting values for $Z/z=1$ are 
 $\epsilon \simeq 5.61$ and $\tilde{c} \simeq 
0.0416$ and are thus very similar to the DH case\cite{comment4}. 
For $Z/z>1$ the
deviations from DH theory are pronounced. In Fig.2 we show coexistence
curves for $Z/z=1$, $2$, and
$10$ (solid, broken, and dotted lines, respectively). 
The coexistence curve as predicted by DH theory
would be indistinguishable from the solid line. 

In summary, we introduced a systematic field theory to describe charged
colloidal suspensions, including fluctuations of and correlations
between charged particles.
Attractive colloidal interactions are predicted for asymmetric 
electrolyte solutions, as realized for multivalent $m:1$ 
salt solutions  and in the neighborhood of charged walls.
A critical point of demixing occurs even in the absence 
of attractive colloidal interactions, whose location in the 
temperature-density plane depends strongly on the colloidal charge.
The hard-core repulsion between colloids and ions has been incorporated
by a small-distance cut-off, similar to 
recent lattice theories\cite{Borukhov}.
This is a rather poor description of colloidal systems, because here a large
size difference between colloids and ions exists. 
We hope to treat size asymmetries more accurately 
in the future using perturbative
treatments of the hard-core interactions between all  particle pairs.

\begin{figure}
\caption{Critical interaction strength $\epsilon \equiv zZ\ell_B/a$
(solid line) and critical volume fraction 
$\tilde{c} \equiv a^3 c(1+Z/z)$ (broken line)
as a function of the colloid/counterion
charge ratio $Z/z$.}
\caption{Two-phase coexistence regions for $Z/z=1$ (solid line),
$2$ (broken line), and $10$ (dotted line) as a function of
volume fraction $\tilde{c}$ and rescaled  temperature $z/Z \epsilon$.}
\end{figure}
\end{multicols}


\begin{references}
\bibitem{comment1}
It is easily seen that a small-distance cutoff $a$ is needed in order
to make the correlation function normalizable. We interpret this
cutoff as an effective hard-core radius.
\bibitem{comment2}
A second interpretation of the DH interaction is
obtained by rewriting  Eq.(\ref{defDH})  in matrix notation as 
$ v_{\rm DH} =  v \left[ \delta + I_2 v \right]^{-1}
 \simeq  v - I_2 v^2 +  I_2^2 v^3 - I_2^3 v^4 + \cdots $.
The DH interaction thus 
is the renormalized interaction which results from all possible
combinations of indirect interaction terms mediated by salt ions.
\bibitem{comment3}
An effective attractive interactions does not necessarily mean
coagulation of particles, since the centrifugal potential (which
is logarithmic) has to be overcome.
\bibitem{comment4}
MC estimates of  critical parameters for $Z/z=1$ range
from $\tilde{c} \simeq 0.07$, $\epsilon \simeq 14.3$\cite{Valleau}
to $\tilde{c} \simeq 0.03$ , $\epsilon \simeq17.5$\cite{Fisher1};
we attribute the discrepancy to our results, especially in
$\epsilon$, to  our approximate way of including the
hardcore repulsion.
The improved  DH theory yields $\tilde{c} \simeq 0.028$ and
$\epsilon \simeq 17.5$ \cite{Fisher1}.
\end{references}
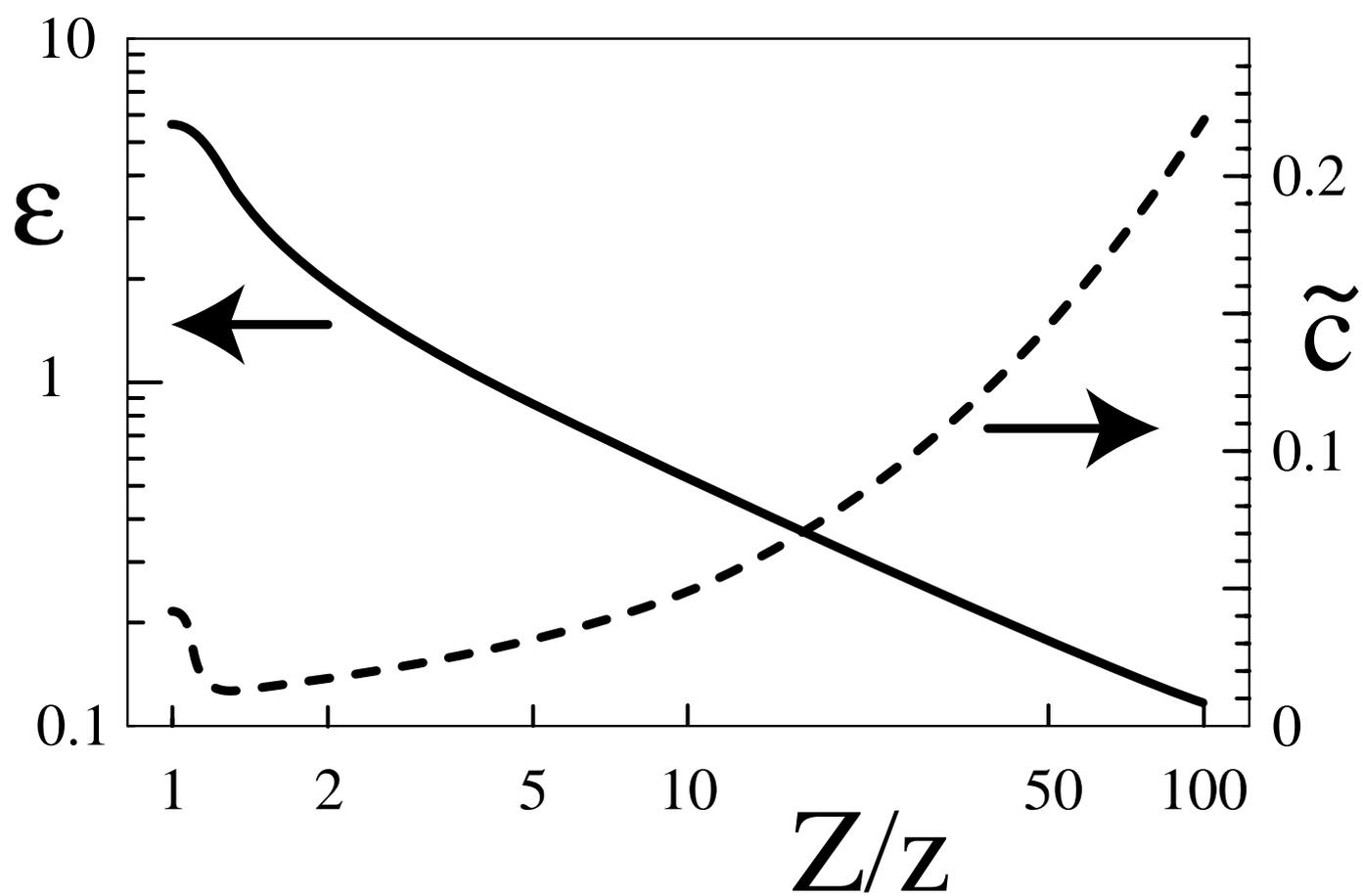
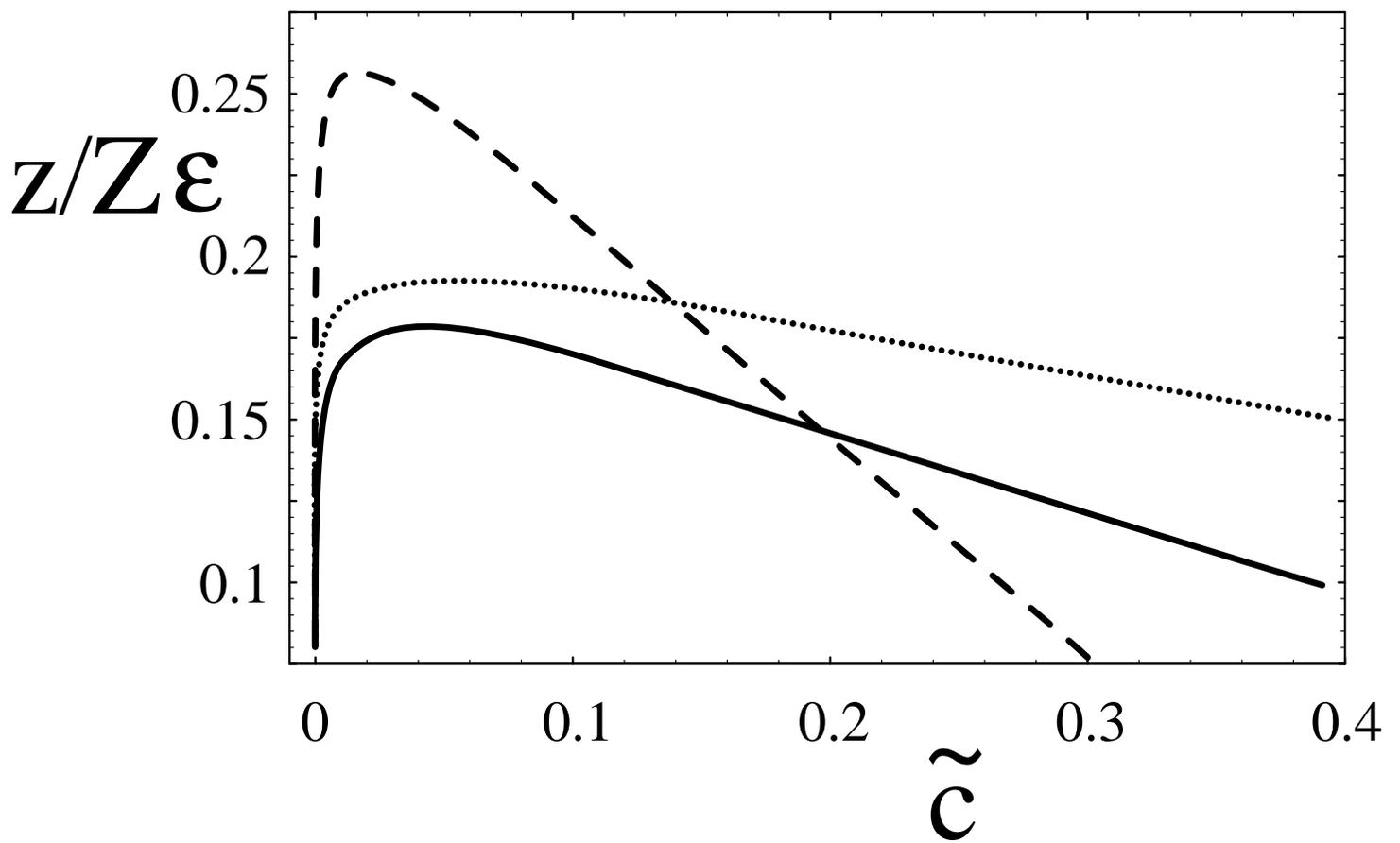
\end{document}